\documentclass[pre,aps,twocolumn,nofootinbib]{revtex4-1}
\usepackage[latin1]{inputenc}
\usepackage[english]{babel}
\usepackage{graphicx}
\usepackage{color}
\usepackage{amsmath}
\usepackage{amssymb}

\begin{document}

\title{\bf Semiclassical corrections to black hole entropy and the generalized uncertainty principle}
%
%
\author{Pedro Bargue\~no}
\email{p.bargueno@uniandes.edu.co}
\affiliation{Departamento de F\'{\i}sica, 
Universidad de los Andes, Apartado A\'ereo {\it 4976}, Bogot\'a, Distrito Capital, Colombia}
\author{Elias C. Vagenas}
\email{elias.vagenas@ku.edu.kw}
\affiliation{Theoretical Physics Group, Department of Physics, Kuwait University,
P.O. Box 5969, Safat 13060, Kuwait}
\begin{abstract}
\par\noindent
In this paper, employing  the path integral method in the framework of a canonical description of a Schwarzschild black hole, we obtain the corrected inverse temperature and entropy of the black hole. The corrections are those coming from the quantum  effects  as well as from the Generalized Uncertainty Principle effects. Furthermore,  an equivalence between the polymer quantization and the  Generalized Uncertainty Principle  description is shown provided the parameters characterizing these two descriptions are proportional.

\end{abstract}

\maketitle 

\section{Introduction}
\par\noindent
After Mead \cite{Mead1964}, who was the first who pointed out the role of gravity on the existence of a fundamental measurable length, a considerably amount of effort has been
devoted to study the modification of the Heisenberg uncertainty principle, known as Generalized Uncertainty Principle (GUP), together with the consequences it leads to
\cite{Amati1989,Maggiore1993,Maggiore1994,Maggiore1993b,
Garay1995,Scardigli1999,Hossenfelder2003,Bambi2008,Hossain2010,Kempf1995,Kempf1997,
Brau1999,Magueijo2002,Magueijo2003,Magueijo2005,Cortes2005,Ghosh2007}. 
A specific form of the GUP and its associated commutation relations, 
together with its physical consequences have
been recently studied \cite{Das2008,Das2009} 
\footnote{It is noteworthy that since there is a plethora of different forms of GUP,  the phenomenological implications of GUP are numerous \cite{Nozari:2010qy,Das:2011tq,Mimasu:2011sa,Pedram:2012dm,Pedram:2013gua,Moussa:2014eda} .}. Moreover, there has been a very  recent interest in a different
version of the GUP \cite{Ali2009,Das2010,Ali2011}, which predicts not only a minimum length but also a maximum
momentum \cite{Magueijo2002,Magueijo2003,Magueijo2005}.

As shown in \cite{Das2008}, the effects of this GUP can be implemented both in classical and quantum systems by defining deformed
commutation relations by means of
\begin{equation}
x_{i}=x_{0i}\; ;  \; p_{i}=p_{0i}\,\left(1-\alpha p_{0}+ 2 \alpha^{2} p_{0}^{2}\right),
\end{equation}
where $\left[x_{0i},p_{0j}\right]=i \hbar \delta_{ij}$ and $p_{0}^{2}=\sum_{j=1}^{3}p_{0j}p_{0j}$ and
$\alpha = \nobreak \alpha_{0}/m_{p}c$, being $\alpha_{0}$ a dimensionless constant. Interestingly, the fact that polymer quantization leads to a modified uncertainty principle
\cite{Hossain2010} has led some authors to think that some forms of GUPs and polymer quantization predict the same physics \cite{Majumder2012}.

Among all quantum gravitational effects one can think of, black hole (BH) entropy can be considered as the paradigmatic one. From the realization that BHs are
thermodynamic objects \cite{Bekenstein1972,Bekenstein1973,Bekenstein1974} which radiate \cite{Hawking1974,Hawking1975}, 
the entropy of a Schwarzschild BH is given by the Bekenstein--Hawking relation
\begin{equation}
S_{BH}= \frac{A_{BH}}{4l_{p}^{2}},
\end{equation}
where $A_{BH}$ is the area of the BH horizon and $l_{p}=\sqrt{\frac{G\hbar}{c^{3}}}$ is the Planck length. After these findings, several approaches to quantum gravity (QG)
have predicted the following form for the QG-corrected BH entropy 
\cite{Mann:1996ze, Mann:1997hm,Kaul2000,Medved2004,Camelia2004,Meissner2004,
Das2002,Domagala2004,Chatterjee2004,Akbar2004,Myung2004,Chatterjee2005}
\begin{equation}
\label{eqfirst}
\frac{S_{QG}}{k}= \frac{A_{BH}}{4 l_{p}^{2}}+c_{0}\ln \left(\frac{A_{BH}}{4 l_{p}^{2}} \right)+\sum_{n=1}^{\infty}c_{n}\left(\frac{A_{BH}}{4l_{p}^{2}} \right)^{-n},
\end{equation}
where the $c_{n}$ coefficients are model--dependent parameters. Specifically, loop quantum gravity calculations are used to fix $c_{0}=-1/2$ \cite{Meissner2004b}.

In particular, the deformed commutation relations previously presented have been widely used to compute the effects of the GUP on the BH entropy from different perspectives 
(see, for example, \cite{Medved2004,Adler2001,Camelia2006,Majumder2011,Majumder2013}), which reads 
\begin{equation}
\frac{S_{GUP}}{k}= \frac{A_{BH}}{4 l_{p}^{2}}+\frac{\sqrt{\pi} \alpha_{0}}{4}\sqrt{\frac{A_{BH}}{4l_{p}^{2}}}-\frac{\pi \alpha_{0}^{2}}{64}\ln \left(\frac{A_{BH}}{4 l_{p}^{2}} \right)+
\mathcal{O}(l_{p}^{3}).
\end{equation}
Therefore, both the logarithmic correction (with the correct sign) and a new term with goes as $\sqrt{A_{BH}}$ can be derived employing GUP.

The paper is organized as follows. In section II, we briefly present how to include quantum  effects on the inverse temperature as well as the entropy  of the Schwarzschild BH  by means of the path integral method which is applied to a canonical description of the BH \cite{Obregon2001}. Using this semiclassical approach, the logarithmic correction in the entropy is also obtained \cite{Obregon2001}. In section III, following the procedure described in section II, GUP effects as well as quantum effects are included \cite{Bina2010}, thus the expressions for the corresponding corrected inverse temperature and entropy of the Schwarzschild BH are obtained. In section IV,  an equivalence between the polymer quantization and the  GUP description is pointed out 
provided that the parameters that characterize these two descriptions are proportional. Finally, in section V, a brief summary of the obtained results is given.

\section{Quantum corrections to black hole entropy}
\par\noindent
Following Ref. \cite{Makela1998},
a Schwarzschild BH can be described as a canonical system by a Hamiltonian that can be identified with its mass, i.e., $H=m$, after an appropriate pair of canonical coordinates $(m,p_{m})$ 
is introduced. However, there is a canonical transformation to a new canonical pair $(a,p_{a})$ such that
the Hamiltonian can now be written as \cite{Makela1998}
\begin{equation}
H= \frac{p_{a}^{2}}{2a}+\frac{a}{2}.
\end{equation}
Performing some computations within the canonical quantization framework, 
the corresponding Wheeler--DeWitt equation can be stated in the form of a 
Schr\"odinger equation for a quantum harmonic oscillator which reads \cite{Obregon2001}
\begin{equation}
\left(-\frac{1}{2}l_{p}^{2}E_{p}\frac{d^{2}}{dx^{2}}+\frac{E_{p}}{2l_{p}^{2}}x^{2}\right)U(x)
=\frac{R_{s}}{4l_{p}}E_{s}U(x),
\label{quantum oscillator}
\end{equation}
where $E_{p}=\sqrt{\frac{\hbar c^{5}}{G}}$ is the Planck energy, $E_{s}= Mc^{2}$ and $R_{s}=\frac{2GM}{c^{2}}$ are the black hole ADM energy\footnote{The black hole ADM mass, i.e., M,  is equal to the mass of the canonical system, i.e., m, when Einstein's equations are satisfied.} and the Schwarzschild radius\footnote{The parameter $s$ is a factor ordering parameter that appears in the quantum equation which has now been transformed into an equation for a quantum linear oscillator, i.e., Eq. (\ref{quantum oscillator}). }, respectively.
The function $U(x)$ is related to the BH wavefunction $\Psi(x)$ by $U(x)=a^{-1} \Psi(x)$, where  the variable $x$ is equal to $(a - R_{s})$.
\par
As shown in Ref. \cite{Obregon2001}, quantum effects on the thermodynamics of the Schwarzschild BH can 
be introduced by means of the path integral method applied to the harmonic oscillator \cite{Feynmanbook}. 
The key point is first to consider the modified harmonic potential given by 
\begin{equation}
V(x)=\frac{m \omega^{2}}{2} \left(x^{2} + \frac{ \beta_{Q}\hbar^{2}}{12 m}\right),
\end{equation} 
where $\omega$ is the frequency of the quantum harmonic oscillator given by $\omega=\sqrt{\frac{3}{2\pi}}\frac{E_{p}}{\hbar}$ and $\beta_{Q}$ is the quantum-corrected inverse BH temperature.
Then, one has to proceed calculating the desired thermodynamic quantities starting from the partition function 
$Z=\nobreak h^{-1}\int_{-\infty}^{\infty} dp \int_{-\infty}^{\infty} dx e^{-\beta H_{Q}}$, where the  
quantum-corrected Hamiltonian is
\begin{equation}
\label{eqHam}
H_{Q}= \frac{p^{2}}{2m_{p}}+\frac{m_{p}\omega^{2}x^{2}}{2} +  \frac{\beta_{Q} E_ {p}^{2}}{16 \pi}, 
\end{equation}
where $m_{p}$ is the Planck mass and is equal to $\sqrt{\frac{\hbar c}{G}}$.
In particular, equating the average (thermodynamical) energy to the internal gravitational energy of the BH, $\bar E= Mc^{2}$, the relation between the quantum-corrected inverse BH temperature and the BH mass can be written as
\begin{equation}
\beta_{Q}= \beta_{H}\left[1-\frac{1}{\beta_{H}M c^{2}}+\mathcal{O}\left(\frac{E_{p}}{Mc^{2}}\right)^{4}\right],
\label{qginversetemp}
\end{equation}
where $\beta_{H}=\frac{8\pi M c^{2}}{E_{p}^{2}}$ is the inverse of Hawking's temperature.
\par\noindent
Furthermore, after defining the BH horizon area, $A_{BH}=4\pi R_{s}^{2}$, the quantum-corrected entropy of the BH can be written as
\begin{equation}
\label{eqS1}
\frac{S_{Q}}{k}=\frac{A_{Q}}{4l_{p}^{2}} - \frac{1}{2}\ln \left(\frac{A_{Q}}{4l_{p}^{2}} \right) - \frac{1}{2}\ln(24) + 1 +\mathcal{O}\left(\frac{E_{p}}{Mc^{2}}\right)^{6},
\end{equation}
where 
\begin{equation}
\label{eqobre}
A_{Q}=A_{BH} \left[1 - \frac{1}{8\pi}\left(\frac{E_{p}}{Mc^{2}}\right)^{2}\right]^{2}
\end{equation}
is a modified BH area which includes quantum corrections \cite{Obregon2001}.
%
%
\section{GUP corrections to quantum black hole entropy}
%
%
%
\par\noindent
By means of the GUP-induced deformed canonical commutator,
a general non--relativistic Hamiltonian of the form $H=\nobreak \frac{p_{0}^{2}}{2m}+V(x)$
transforms into\footnote{Here we are only considering 1D systems.}
\begin{equation}
\label{hamvagenas}
H_{GUP}=\frac{p_{0}^{2}}{2m}+V(x)-\frac{\alpha}{m}p_{0}^{3} + \frac{5\alpha^{2}}{2m}p_{0}^{4} +\mathcal{O}(\alpha^{3}).
\end{equation}
\par\noindent
For the moment, only the quadratic GUP modification will be considered. 
Thus, the  Hamiltonian  of the  Schwarzschild BH which includes GUP and 
quantum corrections now reads
\begin{equation}
\label{eqHam}
H_{GUP+Q}= \frac{p^{2}}{2m_{p}}+\frac{3 m_{p}E_{p}^{2}x^{2}}{4\pi \hbar^{2}}+\frac{5\alpha^{2}}{2m_{p}}p^{4}+\frac{\beta E_ {p}^{2}}{16 \pi},
\end{equation}
where $p$ stands for $p_{0}$ to simplify the notation.
The partition function can be easily calculated giving
\begin{eqnarray}
\label{ZG}
Z_{GUP+Q}&=&\frac{1}{2\alpha E_{p} \sqrt{15  \beta m_{p}}}e^{\frac{\beta_{Q}(\pi-5\alpha^{2}\beta E_{p}^{2}m_{p})}{80\alpha^{2}m_{p}\pi}}K_{1/4}\left(\frac{\beta}{80 \alpha^{2}m_{p}}\right) \nonumber \\
&=&\sqrt{\frac{2\pi}{3}}\frac{1}{\beta E_{p}}e^{-\frac{\beta^{2}E_{p}^{2}}{16\pi}} (1-\frac{15 m_{p}}{2\beta}\alpha^{2} + \mathcal{O}(\alpha^{4})),
\end{eqnarray}
where $K_{i}(x)$ stands for the second modified Bessel function of order $i$.

The internal energy and the entropy can be computed from the standard formulae $\bar E=\nobreak-\frac{\partial \ln(Z)}{\partial\beta}$ and 
$S=\nobreak k(\ln Z) -\nobreak \beta \frac{\partial \ln(Z)}{\partial\beta}$ and, thus, the results will be of the form
\begin{eqnarray}
\bar E_{GUP+Q}&=&\frac{1}{32}\left[\frac{24}{\beta}-\frac{2}{5\alpha^{2}m_{p}}+\frac{4\beta E_{p}^{2}}{\pi}+\frac{
2 K_{3/4}(\frac{\beta}{80\alpha^{2}m_{p}})}
{K_{1/4}(\frac{\beta}{80\alpha^{2}m_{p}})}\right] \nonumber \\
&=&\frac{1}{\beta}+\frac{\beta E_{p}^{2}}{8\pi}-\frac{15 m_{p}}{2\beta^{2}}\alpha^{2}+ \mathcal{O}(\alpha^{4}) 
\label{gup+qgenergy}
\end{eqnarray}
and
\begin{eqnarray}
\label{eqsgup}
\frac{S_{GUP+Q}}{k}&=&\frac{\beta }{32}\left[\frac{24}{\beta}-\frac{5}{2\alpha^{2}m_{p}}+\frac{4\beta E_{p}^{2}}{\pi}+ \frac{2 K_{3/4}(\frac{\beta}{80\alpha^{2}m_{p}})}{K_{1/4}(\frac{\beta}{80\alpha^{2}m_{p}})} \right] \nonumber \\
&+& \ln \left[\frac{1}{E_{p}\sqrt{6 \alpha \beta m_{p}}}e^{\frac{\beta(\pi-5\alpha^{2}\beta E_{p}^{2}m_{p})}{80\alpha^{2}m_{p}\pi}} \right] \\
&  \hspace{-0.5ex}  = \hspace{-0.5ex} &  \hspace{-0.5ex} 1+\frac{\beta^{2}E_{p}^{2}}{16\pi}-\ln \hspace{-0.5ex} \left(\sqrt{\frac{3}{2\pi}}\beta E_{p}\right)
-\frac{15m_{p}}{\beta}\alpha^{2}+ \mathcal{O}(\alpha^{4}) .\nonumber
\end{eqnarray}
As in the previous section, if $\bar E=  Mc^{2}$ is imposed, we obtain, up to first order in $\alpha^{2}$ 
\begin{equation}
\label{eqHaw}
\beta^{3} - \beta_{H}\beta^{2} + \frac{\beta_{H}}{Mc^{2}}\beta=\frac{15 \beta_{H}m_{p}}{2Mc^{2}}\alpha^{2}.
\end{equation}
Therefore, if  the GUP correction as well as quantum corrections are taken into account, the inverse BH temperature reads
\begin{equation}
\label{modift}
\beta_{GUP+Q}= \beta_{H}\left[1 - \frac{1}{\beta_{H}Mc^{2}} + \frac{15 m_{p}}{2\beta_{H}^{2}Mc^{2}}\alpha^{2}+\mathcal{O}\left(\frac{E_{p}}{Mc^{2}}\right)^{4} \right].
\end{equation}
\par\noindent
At this point a couple of comments are in order. First, it is easily seen that employing Eq. (\ref{modift}) and removing the GUP-correction, i.e., $\alpha=0$, the quantum-corrected inverse BH temperature, namely Eq. (\ref{qginversetemp}),  is obtained. 
\par\noindent
Second, employing Eq. (\ref{eqsgup}) and removing the GUP-correction, i.e., $\alpha=0$,  
the corresponding quantum-corrected entropy will now read
\begin{equation}
\frac{S_{Q}}{k}=\frac{A_{Q}}{4l_{p}^{2}}-\frac{1}{2}\ln \left(\frac{A_{Q}}{4l_{p}^{2}} \right)-\frac{1}{2}\ln(24)+1,
\end{equation}
where $A_{Q}$ is the modified BH area given by Eq. (\ref{eqobre}).
%
%
%
\\
\par\noindent
After a long but straightforward calculation, the GUP-corrected quantum entropy  given by Eq. (\ref{eqsgup}) can be written as 
%
%
\begin{widetext}
\begin{equation}
\label{eqspit}
\frac{S_{GUP+Q}}{k}=\frac{A_{GUP+Q}}{4l_{p}^{2}}-\frac{1}{2}\ln \left(\frac{A_{GUP+Q}}{4l_{p}^{2}} \right)-\frac{1}{2}\ln(24)+1
+\frac{15}{2\sqrt{\pi}}\alpha_{0}^{2}\left(\frac{l_{p}^{2}}{A_{BH}}\right)^{1/2}
\end{equation}
\end{widetext}
where the GUP-corrected quantum BH area is of the form
\begin{widetext}
\begin{eqnarray}
\label{eqarea}
A_{GUP+Q}&=&A_{BH} \left[1 - \frac{1}{8\pi}\left(\frac{E_{p}}{Mc^{2}}\right)^{2}+
\frac{15 \pi\alpha_{0}^{2}}{2}\left(\frac{l_{p}^{2}}{A_{BH}}\right)^{3/2}\right]^{2} \nonumber \\
&=&A_{BH} \left[1 - \frac{2 l_{p}^{2}}{A_{BH}}+
\frac{15 \pi\alpha_{0}^{2}}{2}\left(\frac{l_{p}^{2}}{A_{BH}}\right)^{3/2}\right]^{2}
\end{eqnarray}
\end{widetext}
From Eqs. (\ref{eqspit}) and (\ref{eqarea}), it can be explicitly seen that the quantum BH entropy is recovered when $\alpha_{0}=0$. Interestingly, the entropy contains a 
GUP-dependent term not only in the usual area and logarithmic terms (see the fifth term on the right-hand side of Eq. (\ref{eqspit})).

\par\noindent
Furthermore, it is easily seen from the second line of Eq. (\ref{eqarea}) that, for $\alpha_0 \approx 10^{-2}$, the GUP-correction can be considered as negligible compared to the quantum-correction and, thus, $S_{GUP+Q}$ is almost equal to $S_{Q}$ (we are referring to the range of values  for which the entropy is proportional to the area). Then, increasing the value of $\alpha_0$, the entropy $S_{GUP+Q}$ becomes bigger than $S_{Q}$, meaning that the GUP-correction is no more negligible compared to the quantum-correction while, when $\alpha_0 $ becomes of the order of unity, the  entropy $S_{GUP+Q}$ becomes almost double of $S_{Q}$. Finally, in the case where $\alpha_0$ is much bigger than the unity but less than $10^{17}$ \cite{Das2008}, one can say that the GUP-correction dominates and the quantum correction is now negligible.
%
%
%
%
%

\section{Equivalence between polymerization and quadratic GUP}
\par\noindent
It was recently pointed out that when the polymerization protocol is applied to a classical Hamiltonian 
$H=\frac {p^{2}}{2m}+V(x)$, the resulting polymer Hamiltonian, $H_{\mu}$,
can be written as  \cite{Gorji2014}
\begin{equation}
H_{\mu}=\frac{1}{m\mu^{2}}\left[1-\cos(\mu p) \right].
\end{equation}
Therefore, in the semiclassical regime, a $\mu$--dependent classical theory can be obtained from the polymerization process. 
By expanding the kinetic term of the classical polymeric Hamiltonian,
\begin{equation}
H_{\mu}=\frac{1}{m\mu^{2}}\left[1-\cos(\mu p) \right]=\frac{p^{2}}{2m}-\frac{\mu^{2}}{24 m}p^{4}+ \mathcal{O}(\mu^{4})~,
\end{equation}
and comparing it with that of Eq. (\ref{hamvagenas}), an equivalence between these descriptions can be shown, provided that 
\begin{equation}
|\mu|^{2}=60|\alpha|^{2}.
\end{equation}
\par\noindent
In addition, in Ref. \cite{Gorji2014} it is shown that, after introducing the non--canonical transformation $(x,p)\rightarrow \left(X=x,P=\frac{2}{\mu}\sin(\frac{\mu p}{2}) \right)$,
the polymer Hamiltonian reads the standard form
\begin{equation}
H_{\mu}=\frac{P^{2}}{2m}+V(X)
\end{equation}
and, thus, all the polymeric effects are summarized in the density of states, which is given by the expression
\begin{equation}
\frac{1}{h}\int_{|P|<\frac{2}{\mu}}\frac{dX dP}{\sqrt{1-(\mu P/2)^{2}}}~.
\end{equation}
\par\noindent
Therefore, the polymeric partition function for the quantum corrected Schwarzschild BH Hamiltonian is written as 
\begin{eqnarray}
\label{Zmu}
Z_{\mu}&=&\frac{1}{h}\int_{-\infty}^{\infty} dX\int_{|P|<\frac{2}{\mu}}dP\frac{e^{-\beta\left(\frac{P^{2}}{2m_{p}}+\frac{m_{p}\omega^{2}X^{2}}{2}+\frac{E_ {p}^{2}\beta}{16 \pi}\right)}
}{\sqrt{1-(\mu P/2)^{2}}} \nonumber \\
&=&\frac{1}{\mu}\sqrt{\frac{2\pi}{3}}\left(\frac{2\pi}{E_{p} m_{p}\beta} \right)I_{0}\left(\frac{\beta m_{p}}{\mu^{2}} \right)e^{-\beta \left(\frac{m_{p}}{\mu^{2}}+\frac{\beta E_{p}^{2}}
{16 \pi} \right)} \nonumber \\
&=&\sqrt{\frac{2\pi}{3}}\frac{e^{-\frac{\beta^{2}E_{p}^{2}}{16\pi}}}{\beta E_{p}}\left[ 1+\frac{m_{p}}{8\beta}\mu^{2} + \mathcal{O}(\mu^{4})\right],
\end{eqnarray}
where $I_{0}(x)$ denotes the first order modified Bessel function. 
If one now compares the  polymer partition functions given by Eq. (\ref{Zmu}) with 
the GUP-modified partition function given by Eq. (\ref{ZG}), then the equivalence between the two descriptions can be shown, up to second order in the deformation parameters, namely $\alpha$ and $\mu$, provided that   $|\mu|^{2}=60|\alpha|^{2}$, as previously stated.

\section{Discussion and conclusions}

\par\noindent
In this paper, we have computed the corrected inverse temperature and the entropy of a Schwarzschild BH  when the quantum  effects  as well as the GUP effects are present. The computation was performed by employing  the path integral method in the framework of a canonical description of the BH. In this semiclassical context, the logarithmic correction in the expression for the corrected entropy of the Schwarzschild BH is also obtained. In the limiting case in which the GUP parameter is zero, i.e., $\alpha=0$,  expressions for the quantum corrected inverse temperature and entropy of the Schwarzschild BH that already exist in the literature are obtained. 
Finally, an equivalence between the polymer quantization and the  GUP description is pointed out under the condition that the parameters characterizing these two descriptions, namely $\mu$ and $\alpha$, respectively, are proportional.

\section{acknowledgments}
\par\noindent
ECV would like to thank Kourosh Nozari and Octavio Obregon for very useful correspondences.


\end{document}